\documentclass[pre,showpacs,superscriptaddress,eqsecnum,twocolumn]{revtex4}

\usepackage{amsmath}
\usepackage{amsfonts}
\usepackage{amssymb}
\usepackage{color}

\usepackage{graphicx}

\def\vec#1{\mathbf{#1}}

\begin{document}

\title{Physical origin of nonequilibrium fluctuation-induced forces in fluids}

\author{T. R. Kirkpatrick}
\email{tedkirkp@umd.edu}
\affiliation{Institute for Physical Science and Technology, University of Maryland, College Park, Maryland 20877, USA}
\affiliation{Department of Physics, University of Maryland, College Park, Maryland 20877, USA}

\author{J. M. Ortiz de Z\'arate}
\affiliation{Departamento de F\'{\i}sica Aplicada I, Facultad de F\'{\i}sica, Universidad Complutense, 28040 Madrid, Spain}

\author{J. V. Sengers}
\affiliation{Institute for Physical Science and Technology, University of Maryland, College Park, Maryland 20877, USA}

\date{\today}

\begin{abstract}
Long-range thermal fluctuations appear in fluids in nonequilibrium states leading to fluctuation-induced Casimir-like forces. Two distinct mechanisms have been identified for the origin of the long-range nonequilibrium fluctuations in fluids subjected to a temperature or concentration gradient. One is a coupling between the heat or mass-diffusion mode with a viscous mode in fluids subjected to a temperature or concentration gradient. Another one is the spatial inhomogeneity of thermal noise in the presence of a gradient. We show that in fluids fluctuation-induced forces arising from mode coupling are several orders of magnitude larger than those from inhomogeneous noise.
\end{abstract}


\maketitle

\section{INTRODUCTION}

Thermal fluctuations in fluids in nonequilibrium steady states (NESS) are large and very long ranged~\cite{DorfmanKirkpatrickSengers}. The nonequilibrium (NE) fluctuations are particularly spectacular in fluids in the presence of a temperature or concentration gradient. They arise from a coupling between the heat-diffusion or mass-diffusion mode and the viscous mode through the convective term in the fluctuating--hydrodynamics equations. The intensity of the NE temperature or concentration fluctuations varies with the wave number $q$ of the fluctuations as ${q}^{-4}$, as predicted theoretically~\cite{KirkpatrickEtAl,LawSengers,LawNieuwoudt,BelitzKirkpatrickVotja} and confirmed experimentally~\cite{LawEtAl,SegreEtAl1,SegreEtAlPRE,Mixtures3,VailatiGiglio1,GiglioNature,VailatiGiglio3,miJCP,TakacsEtAl2,CerbinoEtAl}. Hence, the NE fluctuations encompass the entire size of the system~\cite{Physica}.

Actually, there are two distinct mechanisms that lead to long-ranged correlations in NESS. One is the mode-coupling mechanism mentioned above. Parenthetically, we note that such mode-coupling terms already play a crucial role in the dynamical properties of equilibrium fluids both near and away from critical points~\cite{DorfmanKirkpatrickSengers}. The second mechanism, commonly considered by other investigators, is one resulting from a spatial dependence of the thermal noise correlations in the presence of a gradient~\cite{RonisProcaccia,Spohn,GarciaEtAl1,MalekMansourEtAl1,PagonabarragaRubi,BreuerPetruccione,SuarezEtAl,GarciaEtAl2,NajafiGolestanian}. In this approach detailed balance is satisfied locally, but not globally~\cite{SuarezEtAl,Derrida2}. In particular, this mechanism has been studied in NESS in the simplest hydrodynamic models with a single conserved quantity. In contrast to thermal fluctuations from mode coupling, the intensity of thermal fluctuations from inhomogeneous noise varies as ${q}^{-2}$. Thus, while indeed being long ranged, these NE fluctuations are much less important than the long-ranged fluctuations from mode coupling.

Earlier two of us~\cite{miStatis} have shown that in the NE structure factor, experimentally accessible by light scattering or shadowgraphy~\cite{BOOK}, the mode-coupling contributions at any wave number ${q}$ are orders of magnitude larger than the contributions resulting from inhomogeneous noise. The purpose of the present paper is to study the relative importance of the two mechanisms for long-range correlations in fluids in NESS in the context of the NE Casimir effect. In some recent papers we have evaluated the mode-coupling contribution to the NE Casimir effect in fluids~\cite{miPRL2,miPRE2014,miCasimirBin}. Subsequently, Animov \textit{et al.}~\cite{AminovEtAl} have reported a study of the NE Casimir effect in a simple diffusion model that only had the mechanism of lack of detailed balance associated with spatial inhomogeneous noise.

In this paper we consider a real fluid system where both mechanisms are operative. In particular, we shall evaluate both types of fluctuations in a one-component fluid in the presence of a temperature gradient. To understand the physical origin of the various terms that may contribute to the NE pressure, we note that in general a temperature gradient can cause normal stresses or pressures, if nonequilibrium thermodynamics is extended to include nonlinear effects. To frame our purpose in a more fundamental context, the pressure we are calculating here corresponds to a nonlinear Onsager-like cross effect causing a NE pressure induced by a temperature gradient:
\begin{equation} \label{GrindEQ__1_1_}
p_{{\rm NE}} =\kappa _{{\rm NL}} \left(\nabla T\right)^{2} .
\end{equation}
Here $\kappa _{{\rm NL}} $ is a kinetic coefficient commonly referred to as a nonlinear Burnett coefficient~\cite{McLennan,WongEtAl}. As has been discussed elsewhere~\cite{PomeauResibois,ErnstDorfman,Brey,Standish}, it is well known that $\kappa _{{\rm NL}}$ diverges linearly in the large system size, $L$, due to long-time-tails effects. To take this divergence into account we wrote in our previous publications~\cite{miPRL2,miPRE2014} $\kappa _{{\rm NL}} $ as $\kappa _{{\rm NL}} =\kappa _{{\rm NL}}^{\left(0\right)} +\kappa _{{\rm NL}}^{\left(1\right)} L$, where $\kappa _{{\rm NL}}^{\left(0\right)}$ is a bare molecular contribution from short-range correlations. More generally, we can include other sub-leading long-range correlations and write $\kappa _{{\rm NL}} $ as
\begin{equation} \label{GrindEQ__1_2_}
\kappa _{{\rm NL}} =\kappa _{{\rm NL}}^{\left(0\right)} +\frac{\kappa _{{\rm NL}}^{\left(-1\right)} }{L} +...+\kappa _{{\rm NL}}^{\left(1\right)} L.
\end{equation}
The ellipsis in Eq. \eqref{GrindEQ__1_2_} indicates terms that vanish faster as $L\to \infty $ than those indicated. Substituting Eq. \eqref{GrindEQ__1_2_} into \eqref{GrindEQ__1_1_}, we may write more generally the expected NE pressure as
\begin{equation} \label{GrindEQ__1_3_}
p_{{\rm NE}} =\kappa _{{\rm NL}}^{\left(0\right)} \left(\nabla T\right)^{2} +\frac{\kappa _{{\rm NL}}^{\left(-1\right)} }{L} \left(\nabla T\right)^{2} +...+\kappa _{{\rm NL}}^{\left(1\right)} L\left(\nabla T\right)^{2} .
\end{equation}
Actually, we shall see that the subleading terms in Eqs. \eqref{GrindEQ__1_2_} and \eqref{GrindEQ__1_3_} also contain logarithmic corrections, $\propto \ln L$.

In our previous publications~\cite{miPRL2,miPRE2014,miCasimirBin} we have shown that mode-coupling effects yield a giant NE fluctuation-induced pressure $p_{{\rm NE}}$ corresponding to the term with coefficient $\kappa _{{\rm NL}}^{\left(1\right)}$ in Eq.~\eqref{GrindEQ__1_3_}. In this paper we show that the breakdown of detailed balance associated with the presence of inhomogeneous thermal noise yields a subleading contribution in Eq. \eqref{GrindEQ__1_3_} proportional to $\propto L^{-1} \ln L$. Thus for large $L$ the mode-coupling contribution is much more important that contributions resulting from the breakdown of detailed balance from spatially inhomogeneous noise. Physically, the mode-coupling term, $\propto L$, arises from NE correlations that grow linearly with the system size $L$, while terms $\propto L^{-1} \ln L$ and $\propto L^{-1} $ arise from NE correlations that decay linearly in space.

We shall proceed as follows. In Section II we specify the relationship between the NE fluctuation-induced Casimir-like pressure and the intensity of the NE temperature fluctuations. In Section III we review the expressions obtained for the NE temperature fluctuations from coupling of hydrodynamic modes in the presence of a temperature gradient. The resulting fluctuation-induced pressures arising from this mode-coupling mechanism are discussed in Section IV.  In Section V we derive the intensity of the NE temperature fluctuations arising from the spatial inhomogeneity of the local-equilibrium correlations of the fluctuating heat flux in the presence of a temperature gradient. We conclude this paper with a comparison between the two types of NE fluctuation-induced pressures in Section VI.  In Section VII we conclude that the mode-coupling contribution to the fluctuation-induced pressures is orders of magnitude more important than contributions from a lack of detailed balance due to the inhomogeneity of thermal noise correlations.

\section{RELATION BETWEEN NE PRESSURE AND NE TEMPERATURE FLUCTUATIONS}

We consider the pressure $p$ as a function of a fluctuating mass density $\rho +\delta \rho$ and a fluctuating energy density $e+\delta e$:
\begin{equation} \label{GrindEQ__2_1_}
p\left(\rho +\delta \rho ,e+\delta e\right)=p\left(\rho ,e\right)+\delta p,
\end{equation}
where $\rho $ and $e$ are the local average mass density and energy density, respectively. We then apply a Taylor expansion up to terms quadratic in $\delta \rho $ and $\delta e$:
\begin{multline} \label{GrindEQ__2_2_}
{p(\rho +\delta \rho ,e+\delta e)=p(\rho ,e)} +\left(\dfrac{\partial p}{\partial \rho } \right)_{e} \delta \rho +\left(\dfrac{\partial p}{\partial e} \right)_{\rho } \delta e \\+\frac{1}{2} \left[\left(\dfrac{\partial ^{2} p}{\partial \rho ^{2} } \right)_{e} \left(\delta \rho \right)^{2} +2\left(\dfrac{\partial ^{2} p}{\partial \rho \partial e} \right)\delta \rho \delta e\right.\\
+\left.\left(\dfrac{\partial ^{2} p}{\partial e^{2} } \right)_{\rho } \left(\delta e\right)^{2} \right].
\end{multline}
The NE enhancement of the temperature fluctuations originates from a coupling of the heat mode with a viscous mode and are unaffected by the sound modes~\cite{KirkpatrickEtAl,LawSengers}, \textit{i.e.,} with vanishing linear pressure fluctuations, so that
\begin{equation} \label{GrindEQ__2_3_}
\left(\frac{\partial p}{\partial \rho } \right)_{e} \delta \rho +\left(\frac{\partial p}{\partial e} \right)_{\rho } \delta e=0,
\end{equation}
and
\begin{equation} \label{GrindEQ__2_4_}
\delta \rho =-\rho \alpha \delta T,
\end{equation}
where $\alpha $ is the thermal expansion coefficient and $\delta T$ is the fluctuation of the local temperature $T$. We then substitute Eqs. \eqref{GrindEQ__2_3_} and \eqref{GrindEQ__2_4_} into Eq. \eqref{GrindEQ__2_2_} and determine the average NE contribution $p_{{\rm NE}} $ to the pressure $p$:
\begin{multline} \label{GrindEQ__2_5_}
p_{{\rm NE}} =\frac{\left(\rho \alpha \right)^{2} }{2} \left[\left(\frac{\partial ^{2} p}{\partial \rho ^{2} } \right)_{e} -2w\left(\frac{\partial ^{2} p}{\partial\rho\partial{e}} \right)\right.\\
+\left.w^{2} \left(\frac{\partial ^{2} p}{\partial e^{2} } \right)_{\rho } \right]\left\langle \left(\delta T\right)^{2} \right\rangle _{{\rm NE}}
\end{multline}
with
\begin{equation} \label{GrindEQ__2_6_}
w=\left(\frac{\partial p}{\partial \rho } \right)_{e} /\left(\frac{\partial p}{\partial e} \right)_{\rho } .
\end{equation}
Only the NE temperature fluctuations $\left\langle \left(\delta T\right)^{2} \right\rangle _{{\rm NE}} $cause a renormalization of the pressure, because the equilibrium temperature fluctuations are already incorporated in the unrenormalized pressure. Taking the pressure as a function of the conserved thermodynamic quantities $\rho $ and $e$ in the expansion \eqref{GrindEQ__2_2_} to identify the NE fluctuation contribution to the pressure can be shown to be consistent with the mechanical definition of the pressure in terms of the microscopic stress tensor in nonequilibrium thermodynamics~\cite{miPRE2014}.

With the aid of some thermodynamic relations~\cite{ErnstHaugeVanLeeuwen2}, Eq. \eqref{GrindEQ__2_5_} can be converted into
\begin{equation} \label{GrindEQ__2_7_}
p_{{\rm NE}} =\frac{\rho c_{p} \left(\gamma -1\right)}{2T} \widetilde{B} \left\langle \left(\delta T\right)^{2} \right\rangle _{{\rm NE}} ,
\end{equation}
where, to shorten notation, we introduced the dimensionless quantity
\begin{equation}
\widetilde{B}= \left[1-\frac{1}{\alpha c_{p} } \left(\frac{\partial c_{p} }{\partial T} \right)_{p}+ \frac{1}{\alpha ^{2} } \left(\frac{\partial \alpha }{\partial T} \right)_{p} \right],
\end{equation}
while $c_{p} $ is the isobaric specific heat capacity and $\gamma$ the ratio of the isobaric and isochoric heat capacities~\cite{miPRE2014}. An alternative expression for the NE fluctuation-induced pressure $p_{{\rm NE}} $ is obtained by noting that $\alpha =-\rho ^{-1} \left(\partial \rho /\partial T\right)_{p} =v^{-1} \left(\partial v/\partial T\right)_{p} $, where $v$ is the specific volume, and using $\left(\partial c_{p} /\partial T\right)_{p} =\left(\partial ^{2} h/\partial T^{2} \right)_{p} $, where $h$ is the specific enthalpy:
\begin{equation} \label{GrindEQ__2_8_}
p_{{\rm NE}} =-\frac{\rho \left(\gamma -1\right)}{2\alpha T} \left[\left(\frac{\partial ^{2} h}{\partial T^{2} } \right)_{p} -\frac{\rho c_{p} }{\alpha } \left(\frac{\partial ^{2} v}{\partial T^{2} } \right)_{p} \right]\left\langle \left(\delta T\right)^{2} \right\rangle _{{\rm NE}} .
\end{equation}
Equation \eqref{GrindEQ__2_8_} is interesting because of its similarity with the expression for the NE pressure induced by concentration fluctuations in a fluid mixture~\cite{miCasimirBin}.

\section{NE TEMPERATURE FLUCTUATIONS FROM COUPLING OF HYDRODYNAMIC MODES}

We consider a fluid layer between two horizontal thermally conducting plates located at $z=0$ and $z=L$ subject to a stationary temperature gradient $\nabla T_{0} $, where $T_{0} \left(z\right)$ is the local average temperature which is a linear function of the coordinate $z$. In this paper the upper plate at $z=L$ has the higher temperature, so that for fluids with a positive thermal expansion coefficient convection is absent for any possible value of the temperature gradient. The temperature fluctuations $\delta T=\delta T\left({\vec r},t\right)$, which depend on the location ${\vec r}$ and the time $t$, satisfy a linearized fluctuating heat equation:
\begin{equation} \label{GrindEQ__3_1_}
\rho c_{p} \left[\frac{\partial \delta T}{\partial t} +\delta {\vec v}\cdot {\boldsymbol\nabla }T_{0} \right]=\lambda \nabla ^{2} \delta T-{\boldsymbol \nabla }\cdot \, \delta {\bf J},
\end{equation}
where $\lambda $ is the thermal conductivity coefficient and where $\delta {\vec J}$ is a fluctuating heat flux~\cite{LawSengers,miChaperRSC2}. This fluctuating heat equation differs from the one in thermal equilibrium by the presence of the term $\delta {\vec v}\cdot {\boldsymbol\nabla }T_{0} $ which causes a coupling of the temperature fluctuations which the velocity fluctuations $\delta {\vec v=}\delta {\vec v}\left({\vec r},t\right)$ which is absent at this linear level in equilibrium. The velocity fluctuations are to be determined from the linearized Stokes equation at constant pressure~\cite{LawSengers,miChaperRSC2}:
\begin{equation} \label{GrindEQ__3_2_}
\rho \frac{\partial \delta {\vec v}{\bf \; }}{\partial t} =\eta \nabla ^{2} \delta {\bf v}-{\boldsymbol\nabla }\cdot \delta {\vec \Pi },
\end{equation}
where $\eta $ is the shear viscosity and $\delta {\vec \Pi }$ a fluctuating stress tensor. In fluctuating hydrodynamics $\delta {\vec J}$ and $\delta {\vec \Pi }$ are assumed to satisfy a local fluctuation-dissipation theorem such that~\cite{LandauLifshitz,SchmitzCohen2,miChaperRSC1}
\begin{equation} \label{GrindEQ__3_3_}
\hspace*{-1em}\left\langle \delta J_{i} \left({\vec r},t\right)\; \delta J_{j} \left({\vec r}',t'\right)\right\rangle =2k_{{\rm B}} T_{0}^{2} \lambda \delta_{ij}~\delta({\vec r}-{\vec r}')~\delta(t-t')
\end{equation}
and
\begin{multline} \label{GrindEQ__3_4_}
\left\langle \delta {\Pi }_{ij} \left({\vec r},t\right)\; \delta {\Pi }_{kl} \left({\vec r}',t'\right)\right\rangle =2k_{{\rm B}} T_{0} \, \eta \left(\delta _{ik} \delta _{jl} +\delta _{il} \delta _{jk} \right)\\
\times \delta \left({\vec r}-{\vec r}'\right)\delta \left(t-t'\right),
\end{multline}
where $k_{{\rm B}} $ is Boltzmann constant.

The fluctuating-hydrodynamics equations \eqref{GrindEQ__3_1_} and \eqref{GrindEQ__3_2_} have been solved in previous publications. In principle all thermophysical properties in these equations depend on temperature and, hence, on the vertical position $z$ in the fluid layer, but in practice we approximate them by their average value in the fluid layer. This approximation has turned out to be in good agreement with NE light-scattering experiments~\cite{SegreEtAl1}. Specifically we approximate $T_{0} \left(z\right)$ in Eqs. \eqref{GrindEQ__3_3_} and \eqref{GrindEQ__3_4_} by the average temperature $\overline{T_{0} }$. The effect of the spatial dependence of $T_0(z)$  on the NE Casimir pressure will be considered in Section~\ref{S5}. The temperature fluctuations should vanish at thermally conducting walls. For the velocity fluctuations both stress-free and rigid boundary conditions have been considered~\cite{Physica,EPJ}. For stress-free boundaries we have been able to obtain an explicit analytic solution~\cite{miPRL2}:
\begin{equation} \label{GrindEQ__3_5_}
\left\langle \left(\delta T\left(z\right)\right)^{2} \right\rangle _{{\rm NE,mc}} =\frac{k_{{\rm B}} \overline{T_{0} }L\left(\nabla T_{0} \right)^{2} }{48\pi \rho D_{T} \left(\nu +D_{T} \right)}~F(z)
\end{equation}
with
\begin{equation} \label{GrindEQ__3_6_}
F(z)=6\frac{z}{L} \left(1-\frac{z}{L} \right),
\end{equation}
where $D_{T} $ is the thermal diffusivity and $\nu =\eta /\rho $ the kinematic viscosity. The subscript NE,mc  indicates that Eq. \eqref{GrindEQ__3_5_} represents the intensity of NE temperature fluctuations arising from a coupling between hydrodynamic modes, \textit{i.e}., from solving the coupled Eqs. \eqref{GrindEQ__3_1_} and \eqref{GrindEQ__3_2_}. The intensity $\left\langle \left(\delta T\right)^{2} \right\rangle _{{\rm NE,mc}} $ of the NE temperature fluctuations depends on the vertical location $z$ in the fluid layer through the function $F$. We note that the average intensity of the NE temperature fluctuations is given by
\begin{equation} \label{GrindEQ__3_7_}
\begin{split}
\overline{\left\langle \left(\delta T\right)^{2} \right\rangle }_{{\rm NE,mc}}& \equiv \frac{1}{L} \int _{0}^{L}dz\left\langle \left(\delta T\right)^{2} \right\rangle _{{\rm NE,mc}}\\
&= \frac{k_{{\rm B}} \overline{T_{0} }\; L\left(\nabla T_{0} \right)^{2} }{48\pi \rho D_{T} \left(\nu +D_{T} \right)} .
\end{split}
\end{equation}

\section{NE FLUCTUATION-INDUCED PRESSURES FROM COUPLING OF HYDRODYNAMIC MODES}

The NE fluctuation-induced pressure $p_{{\rm NE,mc}} $ is obtained by substituting Eq. \eqref{GrindEQ__3_5_} into Eq. \eqref{GrindEQ__2_7_}:
\begin{equation} \label{GrindEQ__4_1_}
p_{{\rm NE,mc}} =\frac{c_{p} k_{{\rm B}} \overline{T_{0} }^{2} \left(\gamma -1\right)}{96\pi D_{T} \left(\nu +D_{T} \right)} \widetilde{B} F\left(z\right)L\left(\frac{\nabla T_{0} }{\overline{T_{0} }} \right)^{2} .
\end{equation}
This fluctuation-induced pressure depends on the position $z$ in the fluid layer. Mechanical equilibrium requires that any induced pressure gradient will cause a rearrangement of the density profile by a NE amount $\rho _{{\rm NE}} \left(z\right)$ so as to create a uniform pressure enhancement $\overline{p}_{{\rm NE}} $. The total pressure is then
\begin{equation} \label{GrindEQ__4_2_}
p=p_{{\rm eq}} +\overline{p}_{{\rm NE}} ,
\end{equation}
where $p_{{\rm eq}} $ is the equilibrium pressure. The fluctuation-induced NE density profile caused by $p_{{\rm NE}} \left(z\right)$ is
\begin{equation} \label{GrindEQ__4_3_}
\rho _{{\rm NE}} \left(z\right)=-\rho \kappa _{T} \left[p_{{\rm NE}} \left(z\right)-\overline{p}_{{\rm NE}} \right],
\end{equation}
where $\kappa _{T} =\rho ^{-1} \left(\partial \rho /\partial p\right)_{T} $ is the isothermal compressibility. Conservation of mass implies that
\begin{equation} \label{GrindEQ__4_4_}
\int _{0}^{L}dz\; \rho _{{\rm NE}} \left(z\right) =-\rho \kappa _{T} \left[\int _{0}^{L}dz\; p_{{\rm NE}} \left(z\right)-L\overline{p}_{{\rm NE}}  \right]=0,
\end{equation}
so that
\begin{equation} \label{GrindEQ__4_5_}
\overline{p}_{{\rm NE}} =\frac{1}{L} \int _{0}^{L}dz\; p_{{\rm NE}} \left(z\right) ,
\end{equation}
which equals the average value of the NE fluctuation-induced pressure in the fluid layer~\cite{miPRL2}. We note that Eqs.~\eqref{GrindEQ__4_3_}~--~\eqref{GrindEQ__4_5_} are independent of the boundary conditions for the fluctuations and independent of the physical origin of the NE temperature fluctuations.

We conclude that the effective uniform NE Casimir-like pressure is obtained by substituting Eq. \eqref{GrindEQ__4_1_} into Eq.~\eqref{GrindEQ__4_5_}:
\begin{equation} \label{GrindEQ__4_6_}
\overline{p}_{{\rm NE,mc}} =\frac{c_{p} k_{{\rm B}} \overline{T_{0} }^{2} \left(\gamma -1\right)}{96\pi D_{T} \left(\nu +D_{T} \right)} \widetilde{B} L\left(\frac{\nabla T_{0} }{\overline{T_{0} }} \right)^{2} .
\end{equation}
It is important to note that for a fixed value of the temperature gradient, the NE fluctuation-induced pressure increases with the distance $L$. Thus we have recovered the contribution in Eq.~\eqref{GrindEQ__1_3_} proportional to $L\left(\nabla T\right)^{2} $ with an explicit expression for the divergent part $\kappa _{{\rm NL}}^{\left(1\right)} $ of the nonlinear Burnett coefficient $\kappa _{{\rm NL}} $

Experimentally, it may be more practical to study the NE fluctuation-induced pressure as a function of the distance $L$ at a fixed temperature difference $\Delta T=L\nabla T_{0} $ between the plates:
\begin{equation} \label{GrindEQ__4_7_}
\overline{p}_{{\rm NE,mc}} =\frac{c_{p} k_{{\rm B}} \overline{T_{0} }^{2} \left(\gamma -1\right)}{96\pi D_{T} \left(\nu +D_{T} \right)} \frac{\widetilde{B}}{L} \left(\frac{\Delta T}{\overline{T_{0} }} \right)^{2} .
\end{equation}
Order of magnitude estimates for the NE fluctuation-induced pressures obtained by substituting the thermophysical properties of liquid water~\cite{IAPWS2011} at 298 K into Eq. \eqref{GrindEQ__4_7_} are presented in Table I. It is interesting to compare the magnitude of the NE fluctuation-induced pressures with the magnitude of the original electromagnetic Casimir pressures $p_{{\rm emf}} $ between two conducting plates~\cite{Casimir49} and the Casimir pressures $p_{{\rm c}} $ induced by critical fluctuations in fluids~\cite{FisherDeGennes}:
\begin{equation} \label{GrindEQ__4_8_}
p_{{\rm emf}} =-\frac{\pi ^{2} }{240} \frac{\hbar c}{L^{4} } ,
\end{equation}
where $\hbar $ is Planck's constant and $c$ is the speed of light~\cite{BordagEtAl}.
\begin{equation} \label{E49}
p_{{\rm c}} =\frac{k_{{\rm B}} T}{L^{3} }~ \Theta \left(L/\xi \right),
\end{equation}
where $\Theta \left(L/\xi \right)$ is a finite-size scaling function with $\xi $ being the correlation length of the critical fluctuations~\cite{GambassiEtAl1}. One commonly defines a universal Casimir amplitude  $\Theta  = {\lim _{x \to 0}}~\Theta(x)$, which however depends on the boundary conditions~\cite{Krech1}. For the 3-dimensional Ising universality class with symmetry-breaking boundary conditions $(+-)$, the experimental value is ${\Theta _{ +  - }} =  + 6 \pm 2$~\cite{FukutoEtAl}. The most recent theoretical estimates are ${\Theta _{ +  - }} =  + 5.42 \pm 0.04$~\cite{VasilyevEtAl} and ${\Theta _{ +  - }} =  + 5.61 \pm 0.02$~\cite{Hasenbusch}.  It is seen from Table~\ref{T1} that, except for distances of the order of $0.1~\mu$m, the NE fluctuation-induced pressures are orders of magnitude larger than either the electromagnetic Casimir pressures or the critical Casimir pressures. We note that $\overline{p}_{{\rm NE,mc}} $ will vary as $L^{-1} $, while $p_{{\rm emf}} $ varies as $L^{-4} $ and $p_{{\rm c}} $ varies as $L^{-3} $ with the distance $L$. Hence, the NE fluctuation-induced pressures $\overline{p}_{{\rm NE,mc}} $ should be observable over a much larger range of distances $L$ than either $p_{{\rm emf}} $ or $p_{{\rm c}} $. And indeed from Table~\ref{T1} we see that ${\overline p _{{\rm{NE,mc}}}}$ at $L = 1~\text{mm}$ becomes already comparable with ${p _{{\rm{emf}}}}$  at $L = 1~\mu$m~\cite{AntoniniEtAl} and at $L = 0.1~\text{mm}$ comparable with ${p _{{\rm{c}}}}$ at $L = 1~\mu$m~\cite{HertleinEtAl,GambassiEtAl2,Gambassi09}. At distances smaller than $0.1~\mu$m, fluctuating hydrodynamics becomes less accurate and other short-range phenomena like van der Waals forces need to be considered.

\begin{table*}
\caption{Estimated Casimir pressures}
\begin{tabular*}{\textwidth}{l@{\extracolsep{\fill}}ccccc}
\toprule
&$L = 10^{-7}$~m&$L = 10^{-6}$~m&$L = 10^{-5}$~m&$L = 10^{-4}$~m&$L = 10^{-3}$~m\\
\colrule
$p_\text{emf}$, Eq.~(4.8)&$-10$~Pa&$-1\times10^{-3}$~Pa	& $-1\times10^{-7}$~Pa	 &$-1\times10^{-11}$~Pa & $-1\times10^{-15}$~Pa \\
$p_\text{c}$,  Eq.~(4.9)\footnote{$\Theta =+5.5$~\cite{VasilyevEtAl,Hasenbusch}.}&$+20$~Pa&$+2\times10^{-2}$~Pa&$+2\times10^{-5}$~Pa&  $+2\times 10^{-8} \; {\rm Pa}$&$+2\times 10^{-11} \; {\rm Pa}$ \\
$\overline{p}_{{\rm NE,mc}} $, Eq.~\eqref{GrindEQ__4_7_}\footnote{Water at $\overline{T_{0} }=298\; {\rm K}$ and $\Delta T=25\; {\rm K}$.}&%
$+10$~Pa & +1 Pa	   & $+1\times10^{-1}$~Pa&$+1\times10^{-2}$~Pa&$+1\times10^{-3}$~Pa\\
$\overline{p}_{{\rm NE,in}}$,  Eq.~(6.6)$^b$&$-1\times10^{-1}$~Pa& $-2\times10^{-4}$~Pa & $-2\times10^{-7}$~Pa&$-3\times10^{-10}$~Pa&$-3\times10^{-13}$~Pa\\
\botrule
\end{tabular*}\label{T1}
\end{table*}

\section{NE TEMPERATURE FLUCTUATIONS FROM INHOMOGENEOUS NOISE\label{S5}}

Recently, an alternative approach for identifying NE fluctuation-induced pressures has been proposed by Aminov \textit{et al}.~\cite{AminovEtAl}. They consider a purely diffusion model in which the NE Casimir forces originate from the local dependence of the correlation function for the random fluctuations of the flux in the presence of a gradient.  As mentioned in the introduction, this mechanism for the appearance of long-range fluctuations in fluids in NESS has earlier been proposed by other investigators~\cite{RonisProcaccia,Spohn,GarciaEtAl1,MalekMansourEtAl1,PagonabarragaRubi,BreuerPetruccione,SuarezEtAl,GarciaEtAl2,NajafiGolestanian}. Aminov \textit{et al}.~\cite{AminovEtAl} determine NE density fluctuations from a fluctuating mass-diffusion equation. For our system this approach requires that we should determine the NE temperature fluctuations from the fluctuating heat-diffusion equation:
\begin{equation} \label{GrindEQ__5_1_}
\rho c_{p} \frac{\partial \delta T}{\partial t} =\lambda \nabla ^{2} \delta T-\boldsymbol\nabla \cdot \delta {\vec J}.
\end{equation}
Equation \eqref{GrindEQ__5_1_} differs from Eq. \eqref{GrindEQ__3_1_} by the absence of a convective term, but the spatial dependence of the amplitude of the correlation function \eqref{GrindEQ__3_3_} for the fluctuating heat flux through the dependence of the local temperature $T_{0} (z)$ on the position $z$ should now be retained:
\begin{equation} \label{GrindEQ__5_2_}
\left\langle \delta J_{i} \left({\vec r},t\right)\; \delta J_{j} \left({\vec r}',t'\right)\right\rangle =2k_{{\rm B}} T_{0}^{2}(z)~\lambda\delta_{ij}~\delta({\vec r}-{\vec r}')~\delta(t-t')
\end{equation}
with
\begin{equation} \label{GrindEQ__5_3_}
T_{0} (z)=\overline{T_{0} }\left[1+\frac{L\nabla T_{0} }{\overline{T_{0} }} \left(\frac{z}{L} -\frac{1}{2} \right)\right].
\end{equation}
Equation \eqref{GrindEQ__5_1_} for the temperature fluctuations has been solved by two of us in a previous publication~\cite{miStatis}. In principle, $\rho $, $c_{p} $, and $\lambda $ in Eq.~\eqref{GrindEQ__5_1_} also depend on the temperature, but it can be readily shown that their dependence on the position $z$ is less important than that of the local temperature $T_{0}$. From Eq.~(17) in Ref.~\cite{miStatis} we find
\begin{multline} \label{GrindEQ__5_4_}
\left\langle \delta T\left({\vec r}\right)\delta T\left({\vec r}'\right)\right\rangle _{{\rm NE,in}} =\frac{k_{{\rm B}} L\left(\nabla T_{0} \right)^{2} }{\rho c_{p} } \int _{0}^{\infty }dq_{\parallel }\\ \times \sum _{N=1}^{\infty }\frac{4\pi q_{\parallel } J_{0} \left(q_{\parallel } r_{\parallel } \right)}{N^{2} \pi ^{2} +q_{\parallel }^{2} L^{2} }   \sin \left(\frac{N\pi z}{L} \right)\sin \left(\frac{N\pi z'}{L} \right),
\end{multline}
where $q_{\parallel } $ is the magnitude of the component of the wave vector $\mathbf{q}$ of the fluctuations in the horizontal \textit{XY} plane and where $J_{0} \left(q_{\parallel } r_{\parallel } \right)$ is a Bessel function with $r_{\parallel}$ being the distance between $\mathbf{r}$ and $\mathbf{r}'$ in the horizontal \textit{XY} plane. The subscript NE,in indicates that these NE fluctuations result from inhomogeneous noise.

For our present purpose we need the intensity of the temperature fluctuations at the same location ${\vec r}={\vec r}'$. A problem is that Eq. \eqref{GrindEQ__5_4_} diverges when ${\vec r}={\vec r}'$~\cite{miStatis}, as was also noticed by Aminov \textit{et al}.~\cite{AminovEtAl}. The physical reason is that fluctuating hydrodynamics ceases to be valid at molecular length scales and we need to separate the fluctuations at long-range hydrodynamic length scales from molecular fluctuations. For a complete theory of NE fluctuations one would need to supplement fluctuating hydrodynamics with kinetic theory for dealing with short-range fluctuations, but that is outside the scope of the present paper.

We find it convenient to introduce a dimensionless integration variable $\widetilde{q}=q_{\parallel } L$. Then
\begin{equation} \label{GrindEQ__5_5_}
\left\langle \left(\delta T\left(z\right)\right)^{2} \right\rangle _{{\rm NE,in}} =\frac{k_{{\rm B}} }{\rho c_{p} } \left(\nabla T_{0} \right)^{2} F_{{\rm NE}} \left(z\right)
\end{equation}
with
\begin{equation} \label{GrindEQ__5_6_}
F_{{\rm NE}} \left(z\right)=\frac{4\pi }{L} \int _{0}^{{\Lambda}L}d\widetilde{q}\; \widetilde{q}\sum _{N=1}^{\infty }\frac{\sin ^{2} \left(N\pi z/L\right)}{N^{2} \pi ^{2} +\widetilde{q}^{2} }   ,
\end{equation}
where we have retained an upper cutoff wave number ${\Lambda}$ corresponding to an inverse microscopic length. To evaluate Eq. \eqref{GrindEQ__5_6_} we first note that~\cite{Physica,Gradstein}
\begin{multline} \label{GrindEQ__5_7_}
\sum _{N=1}^{\infty }\frac{\sin ^{2} \left(N\pi z/L\right)}{N^{2} \pi ^{2} +\widetilde{q}^{2} }\\  =\frac{1}{4\pi \widetilde{q}} \frac{\cosh \left(\widetilde{q}\right)-\cosh \left[\widetilde{q}\left(1-2z/L\right)\right]}{\sinh \left(\widetilde{q}\right)}
\end{multline}
to be substituted into Eq. \eqref{GrindEQ__5_6_}. Physically we should not only retain a molecular cutoff in the integral over the wave numbers but also in the summation so that $N\le {\Lambda }L/2\pi $. We shall take care of this limitation by never letting $z$ or $L-z$ to become microscopically small. As was shown in Section IV, the \textit{z }dependence of the NE temperature fluctuations induces a pressure gradient $dp_{{\rm NE}} \left(z\right)/dz$ that causes a NE contribution $\rho _{{\rm NE}} \left(z\right)$ to the density profile. In the present case this pressure gradient will be determined by $dF_{{\rm NE}} /dz$. With a finite cutoff we can interchange differentiation and integration, so that
\begin{equation} \label{GrindEQ__5_8_}
\frac{dF_{{\rm NE}} }{dz} =-\frac{2}{L^{2} } \int _{0}^{{\Lambda}L}d\widetilde{q} \; \widetilde{q}~\frac{\sinh \left[\widetilde{q}\left(1-2z/L\right)\right]}{\sinh \left(\widetilde{q}\right)} .
\end{equation}
Taking the limit ${\Lambda \; }\to \infty $ we obtain~\cite{Gradstein}
\begin{equation} \label{GrindEQ__5_9_}
\frac{dF_{{\rm NE}} }{dz} =-\frac{1}{2L^{2} } \left[\zeta \left(2,1-z/L\right)-\zeta(2,z/L\right]
\end{equation}
with the generalized Riemann zeta function
\begin{equation} \label{GrindEQ__5_10_}
\zeta\left(2,x\right)=\sum _{N=0}^{\infty }\frac{1}{\left(N+x\right)^{2} }.
\end{equation}
Separating out the $N=0$ term in Eq. \eqref{GrindEQ__5_10_}, we rewrite Eq. \eqref{GrindEQ__5_9_} as
\begin{multline} \label{GrindEQ__5_11_}
\frac{dF_{{\rm NE}} }{dz} =\left[\frac{1}{z^{2} } -\frac{1}{\left(L-z\right)^{2} } \right]\\+\frac{1}{2L^{2} } \left[\widetilde{\zeta}\left(2,z/L\right)-\widetilde{\zeta}\left(2,1-z/L\right)\right]
\end{multline}
with
\begin{equation} \label{GrindEQ__5_12_}
\widetilde{\zeta}\left(2,x\right)=\sum _{N=1}^{\infty }\frac{1}{\left(N+x\right)^{2} }  .
\end{equation}
To obtain $F_{{\rm NE}} \left(z\right)$ we integrate Eq. \eqref{GrindEQ__5_11_} subject to the appropriate boundary conditions for the fluctuations. Our description in terms of fluctuating hydrodynamics ceases to be valid at microscopic distances from the walls and will only be valid in the interval
\begin{equation} \label{GrindEQ__5_13_}
n\sigma \le z\le L-n\sigma ,
\end{equation}
where $\sigma $ represents a molecular size and where $n\gg 1$. We thus require that
\begin{equation} \label{GrindEQ__5_14_}
F_{{\rm NE}} \left(n\sigma \right)=F_{{\rm NE}} (L-n\sigma )=0.
\end{equation}
We then obtain
\begin{multline} \label{GrindEQ__5_15_}
F_{{\rm NE}} \left(z\right)=\frac{1}{2} \left[\frac{1}{n\sigma } -\frac{1}{z} +\frac{1}{L-n\sigma } -\frac{1}{L-z} \right]\\+\frac{1}{2L} \int _{0}^{z/L}dx\left[\widetilde{\zeta }\left(2,x\right)-\widetilde{\zeta}\left(2,1-x\right)\right]
\end{multline}
valid in the range given by Eq. \eqref{GrindEQ__5_14_}. Differentiation of Eq. \eqref{GrindEQ__5_15_} indeed reproduces Eq. \eqref{GrindEQ__5_11_}, while the boundary condition \eqref{GrindEQ__5_13_} is satisfied if we neglect terms that vanish as $n\sigma /L\to 0$. We find it convenient to rewrite Eq. \eqref{GrindEQ__5_15_} as
\begin{multline} \label{GrindEQ__5_16_}
F_{{\rm NE}} \left(z\right)=\frac{1}{2} \left[\frac{1}{n\sigma } -\frac{1}{z} +\frac{1}{L-n\sigma } -\frac{1}{L-z} \right]\\+\frac{1}{2L} \left[\int _{0}^{z/L}dx-\int _{1-z/L}^{1}dx  \right]\widetilde{\zeta}\left(2,x\right).
\end{multline}
Of particular interest in Eq.~\eqref{GrindEQ__5_16_} are the terms $1/z$ and $1/\left(L-z\right)$. From Eq.~\eqref{GrindEQ__5_6_} we see that the correlation function varies in wavenumber space as $1/q^{2} $, which would imply a $1/r$ decay in real space, if two different space points were considered. For evaluating the NE pressure we need the correlation function at a single space point. The walls at $z=0$ and $z=L$ break the translational symmetry. This implies that we should find terms that slowly decay as $1/z$ and $1/\left(L-z\right)$ as one moves away from the two walls.

\section{NE FLUCTUATION-INDUCED PRESSURES FROM INHOMOGENEOUS NOISE\label{S6}}

If we substitute Eq. \eqref{GrindEQ__5_5_} into Eq. \eqref{GrindEQ__2_7_}, we obtain for the NE pressure enhancement:
\begin{equation} \label{GrindEQ__6_1_}
p_{{\rm NE.in}} \left(z\right)=\frac{k_{{\rm B}} \overline{T_{0} }\left(\gamma -1\right)}{2} \widetilde{B} \left(\frac{\nabla T_{0} }{\overline{T_{0} }} \right)^{2} F_\text{NE} \left(z\right).
\end{equation}
As discussed in Section IV, to obtain the effective NE pressure we need to take the spatial average of Eq. \eqref{GrindEQ__6_1_} in accordance with Eq.~\eqref{GrindEQ__4_5_}:
\begin{equation} \label{GrindEQ__6_2_}
\overline{p}_{{\rm NE.in}} =\frac{k_{{\rm B}} \overline{T_{0} }\left(\gamma -1\right)}{2} \widetilde{B} \left(\frac{\nabla T_{0} }{\overline{T_{0} }} \right)^{2} \overline{F}_\text{NE}
\end{equation}
with
\begin{equation} \label{GrindEQ__6_3_}
\begin{split}
\overline{F}_\text{NE}& =\frac{1}{L} \int _{n\sigma }^{L-n\sigma }dz\; F_{{\rm NE}}  \left(z\right)\\&=\frac{1}{2} \left[\frac{1}{n\sigma } -\frac{1}{L} -\frac{2}{L} \ln \left(\frac{L}{n\sigma } \right)\right.\\
&\hspace*{6.1em}-\left.\frac{1}{L} \int _{0}^{1}dx\; \left(1-2x\right)\widetilde{\zeta}\left(2,x\right) \right],
\end{split}
\end{equation}
where we have again neglected some terms that vanish as $n\sigma /L\to 0$. The first term in Eq. \eqref{GrindEQ__6_3_} is a molecular contribution independent of $L$. Upon substituting it into Eq. \eqref{GrindEQ__6_2_} it gives a contribution to the bare NE pressure with coefficient $\kappa _{{\rm NL}}^{\left(0\right)}$ in Eq. \eqref{GrindEQ__1_3_}. Kinetic theory should be able to give the actual finite molecular contribution to the bare coefficient $\kappa _{{\rm NL}}^{\left(0\right)} $. The first two terms in Eq.~\eqref{GrindEQ__6_3_} arise from the terms $1/n\sigma $ and $1/\left(L-n\sigma \right)$ of short-range molecular origin in Eq.~\eqref{GrindEQ__5_16_}; they are independent of $z$ and therefore do not yield any contribution to a NE density profile, \textit{i.e.}, they cancel when the difference is taken between $p_\text{NE}(z)$ and $\overline{p}_\text{NE}$ in Eq.~\eqref{GrindEQ__4_3_}. The terms $1/z$ and $1/\left(L-z\right)$ in Eq.~\eqref{GrindEQ__5_16_} account for long-range fluctuations at hydrodynamic length scales. They will cause a NE density profile in the fluid just as the critical fluctuations responsible for the critical Casimir effect lead to a density or concentration profile~\cite{Krech1,HertleinEtAl,GambassiEtAl2,KrechBook}. Retaining only terms that arise from the long-range correlations, we identify the NE fluctuation induced pressure as
\begin{equation} \label{GrindEQ__6_4_}
\overline{p}_{{\rm NE,in}} =\frac{k_{{\rm B}} \overline{T_{0} }\left(\gamma -1\right)}{2} \frac{\widetilde{B}}{L} \left(\frac{\nabla T_{0} }{\overline{T_{0} }} \right)^{2} \overline{G},
\end{equation}
where $\overline{G}$ is a dimensionless quantity:
\begin{multline} \label{GrindEQ__6_5_}
\overline{G}=L\overline{F}_{{\rm NE}} =-\ln \left(\frac{L}{n\sigma } \right)-\frac{1}{2} \int _{0}^{1}dx\; \left(1-2x\right) \widetilde{\zeta}\left(2,x\right)\\\simeq -\ln \left(\frac{L}{n\sigma } \right)-0.077\simeq -\ln \left(\frac{L}{n\sigma } \right).
\end{multline}
In the last approximate equality in Eq.~\eqref{GrindEQ__6_3_} we have used that as $L\to \infty $, \textit{i.e.,} for macroscopic values of $L$, the leading term is the logarithmic one. From Eqs.~\eqref{GrindEQ__6_4_} and~\eqref{GrindEQ__6_5_} we see that that the NE pressure arising from noise inhomogeneity gives a contribution to the term in Eq.~\eqref{GrindEQ__1_3_} with coefficient $\kappa _{{\rm NL}}^{\left(-1\right)} $ with a logarithmic factor.

As discussed in Section IV, in practice one may want to study this fluctuation-induced force as a function of $L$ at a fixed temperature difference $\Delta T=L\nabla T_{0} $ between the plates:
\begin{equation} \label{GrindEQ__6_6_}
\overline{p}_{{\rm NE,in}} =\frac{k_{{\rm B}} \overline{T_{0} }\left(\gamma -1\right)}{2} \frac{\widetilde{B}}{L^{3} } \left(\frac{\Delta T_{0} }{\overline{T_{0} }} \right)^{2} \overline{G}.
\end{equation}
Unlike the NE pressure \eqref{GrindEQ__4_7_} from mode coupling, the NE pressure \eqref{GrindEQ__6_6_} from inhomogeneous noise depends on a molecular cutoff through the term $\ln \left(L/n\sigma \right)$ in Eq.~\eqref{GrindEQ__6_5_} for $\overline{G}\left(z\right)$. To get some order-of-magnitude estimates for $\overline{p}_{{\rm NE,in}}$ we shall in practice approximate the cutoff length by the cube root of the molecular volume $v_{0}$:
\begin{equation} \label{GrindEQ__6_7_}
n\sigma \simeq \left(v_{0} \right)^{1/3} .
\end{equation}
Estimated values thus obtained for $\overline{p}_{{\rm NE,in}}$ in liquid water at $\overline{T_{0} }=298\; {\rm K}$ and $\Delta T=25\; {\rm K}$ are included in Table~\ref{T1}. The NE Casimir pressures from inhomogeneous noise are orders of magnitude smaller than the NE Casimir pressures from mode coupling. The physical reason is that in the absence of boundaries, fluctuations from inhomogeneous noise vary with the wave number $q$ as $q^{-2}$ like critical fluctuations, while fluctuations from mode coupling vary as $q^{-4} $. Thus in fluids NE fluctuations from inhomogeneous noise will always be negligible compared to those from mode coupling.

\section{CONCLUSION}

Long-range NE correlations in fluids cause NE Casimir effects. First, they establish a NE contribution to the density profile in the fluid. Second, they induce a NE contribution to the pressure. In principle there are two mechanisms that may cause NE fluctuation-induced pressures in fluids, namely, coupling between hydrodynamic modes and inhomogeneous thermal noise. Considering a fluid subjected to a temperature gradient $\nabla T_{0} $ as an example, we have shown that NE Casimir pressures from mode coupling increase with the distance $L$ as $L\left(\nabla T_{0} \right)^{2}$, while those from inhomogeneous noise decrease as $L^{-1} \ln \left(L\right)\; \left(\nabla T_{0} \right)^{2} $. As a consequence, NE fluctuation-induced forces from mode coupling are orders of magnitude larger than those from inhomogeneous noise.

\begin{acknowledgments}

The authors acknowledge valuable discussions with J.R. Dorfman. The research at the University of Maryland was supported by the U.S. National Science Foundation under Grant No. DMR-1401449. The research at UCM was funded by the Spanish State Secretary of Research under Grant No. FIS2014-58950-C2-2-P. The authors thank one of the referees for some useful suggestions concerning the comparison with critical Casimir pressures.

\end{acknowledgments}


\end{document}